\documentstyle[preprint,prd,aps,psfig,floats]{revtex}

\draft

\begin{document}

\tighten

\title{
\hbox to\hsize{\large Submitted to Phys.~Rev.~D.\hfil E-Print
astro-ph/9809242}
\vskip1.55cm
Probing Grand Unified Theories with Cosmic Ray,\\
Gamma-Ray and Neutrino Astrophysics}

\author{G\"unter Sigl and Sangjin Lee}
\address{Department of Astronomy \& Astrophysics,
Enrico Fermi Institute, The University of Chicago, Chicago, IL~~60637-1433}

\author{Pijushpani Bhattacharjee}
\address{Laboratory for High Energy Astrophysics, Code 661, 
NASA/Goddard Space Flight Center, Greenbelt, Maryland 20771\\
and\\
Indian Institute of Astrophysics, Bangalore-560 034. India.}

\author{Shigeru Yoshida}
\address{Institute for Cosmic Ray Research, University of Tokyo,
Tanashi, Tokyo 188-8502, Japan}

\maketitle

\begin{abstract}
We explore scenarios where the highest energy cosmic rays
are produced by new particle physics near the grand unification
scale. Using detailed numerical
simulations of extragalactic nucleon, $\gamma$-ray, and neutrino
propagation, we show the existence of an interesting
parameter range for which such scenarios may explain part
of the data and are consistent with
all observational constraints. A combination of proposed
observatories for ultra-high energy cosmic rays, neutrino
telescopes of $\gtrsim$ few kilometer scale, and $\gamma$-ray
astrophysics instruments should be able to test
these scenarios. In particular, for neutrino masses in the
eV range, exclusive neutrino decay
modes of superheavy particles can give rise to neutrino fluxes
comparable to those predicted in models of active galactic nuclei.

\end{abstract}


\narrowtext

\section{Introduction}
The highest energy cosmic ray (HECR) events observed above
$100\,$EeV (1 EeV$=10^{18}$ eV)~\cite{fe1,agasa1} are difficult
to explain within conventional models involving first order Fermi
acceleration of charged particles at astrophysical
shocks~\cite{Blandford}. It is hard to accelerate 
protons and heavy nuclei up to such energies even in the most
powerful astrophysical objects~\cite{Hillas} such as radio
galaxies and active galactic nuclei. Also, nucleons above
$\simeq70\,$EeV lose energy drastically due to 
photo-pion production on the cosmic microwave background (CMB) 
--- the Greisen-Zatsepin-Kuzmin (GZK) effect~\cite{GZK} --- 
which limits the distance to possible sources to less than
$\simeq100\,$Mpc~\cite{SSB}. Heavy nuclei are photodisintegrated
in the CMB within a few Mpc~\cite{Puget}. There are no 
obvious astronomical sources within $\simeq 100$ Mpc of the
Earth.

A way around these difficulties is to suppose the HECR are created
directly at energies comparable to or exceeding the observed ones rather
than being accelerated from lower energies. In the current
versions of such ``top-down'' (TD) scenarios, predominantly
$\gamma$-rays and neutrinos are initially produced at ultra-high
energies (UHEs) by the 
decay of supermassive elementary ``X" particles related
to some grand unified theory (GUT). Such X particles could be
released from topological defect relics of phase
transitions which might have been caused by spontaneous breaking
of GUT symmetries in the early Universe~\cite{BHS}. 
TD models of this type are attractive 
because they predict injection spectra
which are considerably harder than shock acceleration spectra
and, unlike the GZK effect for nucleons, there is no threshold
effect in the attenuation of UHE $\gamma$-rays.

There has been considerable discussion in the literature whether
the $\gamma$-ray, nucleon, and neutrino fluxes predicted by TD
scenarios are consistent with all the relevant observational data and
constraints at various energies~\cite{Chi,SJSB,PJ,PS,SLSC}.
The absolute flux levels predicted by TD models are in general
uncertain. While some (though perhaps not all) processes involving
cosmic strings seem to yield negligibly low fluxes~\cite{GK}, other
processes such as those involving annihilation of 
magnetic monopole-antimonopole pairs~\cite{Hill,BS},
cosmic necklaces~\cite{BV}, and possible~\cite{Hindmarsh} (but currently
controversial~\cite{Shellard}) direct emission of X particles
from cosmic strings~\cite{BSS,WMB} can, for reasonable values of
parameters, yield X
particles at rates sufficient to explain the observed HECR flux. 

In this work, instead of trying to calculate the absolute fluxes
in specific TD models, we use the strategy to numerically
calculate the fluxes of nucleons, $\gamma$-rays, and neutrinos,
``optimally'' normalize them to match data and constraints, and
discuss the feasibility and consequences of a set of most ``favorable''
ranges of the relevant parameters implied by our calculations. 

A major new feature of our calculations is that our ``all particle''
propagation code includes the feed-back effect of neutrino cascading on
the electromagnetic and hadronic channels in a fully self-consistent
manner (see below). In addition, spurred by recent experimental
indications of a possible small neutrino mass, we include in our
calculations the effects of a small neutrino mass ($\sim$ eV) and the
consequent Z-boson resonance in the interaction of UHE neutrinos with the
thermal neutrinos.

\section{Top-Down Models}
The X particles released from topological defects 
could be gauge bosons, Higgs bosons, superheavy fermions,
etc.~depending on the specific GUT. These X particles would have
a mass $m_X$ comparable to the symmetry breaking scale and would
rapidly decay into leptons and/or quarks of roughly
comparable energy. We will accordingly consider several possibilities
for the decay products. Prior calculations were restricted
to decay into only quarks. The quarks interact strongly and 
hadronize into nucleons ($N$s) and pions, the latter
decaying in turn into $\gamma$-rays, electrons, and neutrinos. 
Given the X particle production rate, $dn_X/dt$, the effective
injection spectrum of particle species $a$ ($a=\gamma,N,e^\pm,\nu$) 
via the hadronic channel can be
written as $(dn_X/dt)(2/m_X)(dN_a/dx)$,
where $x \equiv 2E/m_X$, and $dN_a/dx$ is the relevant
fragmentation function (FF). For the total hadronic FF, $dN_h/dx$, we use
solutions of the QCD evolution equations in modified leading
logarithmic approximation which provide good fits to accelerator
data at LEP energies~\cite{detal}, as well as recently suggested
extensions for supersymmetry~\cite{BK} (we abbreviate these cases
by ``no-SUSY'' and ``SUSY'', respectively). The difference
in the results for these two choices will be a measure of the
uncertainty associated with the FF. Furthermore, the nucleon content
$f_N$ of the hadrons is
assumed to be in the range 3 to 10\%, and the rest pions distributed
equally among the three charge states (see, however, Ref.~\cite{BirSar}).
The standard pion decay spectra then
give the injection spectra of $\gamma$-rays, electrons, and
neutrinos. The X particle injection rate is assumed to be
spatially uniform and in the matter-dominated era can be
parametrized as $dn_X/dt\propto t^{-4+p}$~\cite{BHS},
where $p$ depends on the specific defect scenario. In this
paper we focus on the case $p=1$ which is representative of a
number of specific TD processes involving ordinary cosmic
strings~\cite{BR,Hindmarsh,BSS,WMB}, necklaces~\cite{BV} and magnetic 
monopoles~\cite{BS}. Finally, we assume that the X particles
are nonrelativistic at decay.

\section{Numerical Simulations}
The $\gamma$-rays and electrons produced by  X particle decay
initiate  electromagnetic
(EM) cascades on low energy radiation fields such as the
CMB. The high energy photons undergo electron-positron pair
production (PP; $\gamma \gamma_b \rightarrow e^- e^+$), and 
at energies below $\sim 10^{14}$ eV they interact mainly with 
the universal infrared and optical (IR/O) backgrounds, while above 
$\sim 100$ EeV  they interact mainly with the universal radio background (URB).
In the Klein-Nishina regime, where the center of mass energy is
large compared to the electron mass, one of the outgoing particles usually
carries most of the initial energy. This ``leading''
electron (positron) in turn can transfer almost all of its energy to
a background photon via inverse
Compton scattering (ICS; $e \gamma_b \rightarrow e' \gamma$).
EM cascades are driven by this cycle of PP and ICS.
The energy degradation of the ``leading'' particle in this cycle
is slow, whereas the total number of particles grows
exponentially with time. This makes a standard Monte Carlo
treatment difficult. We have therefore used 
an implicit numerical scheme to solve the relevant kinetic 
equations. A detailed account of our transport equation approach
is in Ref.~\cite{Lee}. We include all
EM interactions that influence the $\gamma$-ray spectrum in the energy range
$10^8\,{\rm eV} < E < 10^{25}\,$eV, namely PP, ICS, triplet pair
production (TPP; $e \gamma_b
\rightarrow e e^- e^+$), and double pair production ($\gamma \gamma_b
\rightarrow e^-e^+e^-e^+$), as well as synchrotron losses
of electrons in the large scale extragalactic magnetic field
(EGMF).

Similarly to photons, UHE neutrinos give rise to neutrino
cascades in the primordial neutrino background via exchange
of W and Z bosons~\cite{weiler1,yoshida}. Besides the secondary
neutrinos which drive the neutrino cascade, the W and Z decay products
include charged leptons and quarks which in turn feed into the
EM and hadronic channels. Neutrino interactions become
especially significant if the relic neutrinos have masses $m_\nu$
in the eV range and thus constitute hot dark matter, because
the Z boson resonance then occurs at an UHE neutrino energy
$E_{\rm res}=4\times10^{21}({\rm eV}/m_\nu)$ eV. In fact, this has been
proposed as a significant source of HECRs~\cite{weiler2,YSL}.
Motivated by recent experimental evidence for neutrino mass
we assumed a mass of 1 eV for all three neutrino flavors and
implemented the relevant W boson interactions in the t-channel
and the Z boson exchange via t- and s-channel. Hot dark matter
is also expected to cluster, potentially increasing secondary
$\gamma$-ray and nucleon production~\cite{weiler2,YSL}. This influences
mostly scenarios where X decays into neutrinos only. We
parametrize massive neutrino clustering by a length scale $l_\nu$
and an overdensity $f_\nu$. Values of $l_\nu\simeq$ few Mpc
and $f_\nu\simeq20$ are conceivable on the local Supercluster 
scale~\cite{YSL}.

The relevant nucleon interactions implemented are
pair production by protons ($p\gamma_b\rightarrow p e^- e^+$),
photoproduction of single or multiple pions ($N\gamma_b \rightarrow N
\;n\pi$, $n\geq1$), and neutron decay. Production of secondary
$\gamma$-rays, electrons, and neutrinos by pion decay is also
included, but is in general negligible in the context of
TD scenarios where injection is dominated by $\gamma$-rays and
neutrinos over nucleons.
We assume a flat Universe with no cosmological constant,
and a Hubble constant of $h=0.65$ in units of
$100\;{\rm km}\;{\rm sec}^{-1}{\rm Mpc}^{-1}$ throughout.
An important difference with respect to past work is that we follow
{\it all} produced particles in the EM, hadronic,
and neutrino channel, whereas the often-used continuous energy loss (CEL)
approximation (e.g., \cite{ABS}) follows
only the leading cascade particles. We find that the CEL approximation
can significantly underestimate the cascade flux at lower energies.

The two major uncertainties in the particle transport are the
intensity and spectrum of the URB for which there exists only
an estimate above a few MHz frequency~\cite{Clark}, and the average value
of the EGMF. To bracket these uncertainties we performed simulations
for the observational URB estimate from Ref.~\cite{Clark} that
has a low-frequency cutoff at 2 MHz (``minimal''), and the medium
and maximal theoretical estimates
from Ref.~\cite{PB}, as well as for EGMFs between zero
and $10^{-9}$ G, the latter motivated by limits from
Faraday rotation measurements~\cite{kronberg}. A strong URB tends
to suppress the UHE $\gamma$-ray flux by direct absorption
whereas a strong EGMF blocks EM cascading (which otherwise develops
efficiently especially in a low URB) by synchrotron cooling
of the electrons.

\section{Particle Fluxes}

\begin{table}[ht]
\caption[...]{\label{tab:T1}
Some viable $p=1$ TD scenarios explaining HECRs at least above 100 EeV.}
\smallskip
\begin{tabular}[9]{ccccccccc}
$m_X/$GeV & Fig. & URB & EGMF$/$G & FF & $f_N$ & mode &
 $\lesssim$ GZK$^b$ & $\gtrsim$ GZK$^b$ \\
\noalign{\vskip3pt\hrule\vskip3pt}
$10^{13}$ & \ref{fig:F4}
 & \multicolumn{4}{c}{$f_\nu l_\nu\gtrsim400$ Mpc for high URB, no EGMF$^a$}
 & $\nu\nu$ & $\gamma$ & $\gamma$ \\
 & \ref{fig:F3} & high & any & no-SUSY & 10\% &
 $qq$ & $N$ & $N$ \\
 & & $\lesssim$ med & $\lesssim10^{-11}$ & no-SUSY & $\lesssim10\%$
 & $qq$ & $N$ & $\gamma$ \\
 & \ref{fig:F3} & high & $\lesssim10^{-11}$ & no-SUSY & 10\% & $ql$ & $N$
 & $\gamma$ \\
 & & $\lesssim$ med & $\lesssim10^{-11}$ & any & $\lesssim10\%$ & $ql$
 & $\gamma$ & $\gamma$ \\
 & & any & $\lesssim10^{-11}$ & -- & -- & $ll$, $l\nu$ & $\gamma$ &
 $\gamma$ \\
\noalign{\vskip3pt\hrule\vskip3pt}
$10^{14}$ & \ref{fig:F4}
 & \multicolumn{4}{c}{$f_\nu l_\nu\gtrsim150$ Mpc for high URB, no EGMF$^a$}
 & $\nu\nu$ & $\gamma$ & $\gamma$ \\
 & & high & any & no-SUSY & 10\% & $qq$ & $N$
 & $\gamma+N$, $N^c$ \\
 & & $\lesssim$ med & $\lesssim10^{-10}$ & no-SUSY & $\lesssim10\%$ &
 $qq$, $q\nu$ & $\gamma+N$ & $\gamma $ \\
 & & any & $\lesssim10^{-11}$ & any & $\lesssim10\%$ & $ql$ &
 $N$ & $\gamma$ \\
 & & any & $\lesssim10^{-11}$ & -- & -- & $ll$, $l\nu$ & $\gamma$ &
 $\gamma$ \\
\noalign{\vskip3pt\hrule\vskip3pt}
$10^{15}$ &
 & \multicolumn{4}{c}{$f_\nu l_\nu\gtrsim500$ Mpc for high URB, no EGMF$^a$}
 & $\nu\nu$ & $\gamma$ & $\gamma$ \\
 & & any & any & any & $10\%$ & $qq$, $ql$, $q\nu$ & $N$ & \\
 & & $\lesssim$ med & $\lesssim10^{-11}$ & any & $\lesssim10\%$
 & $qq$, $ql$, $q\nu$ & & \\
 & & any & $\lesssim10^{-11}$ & -- & -- & $ll$, $l\nu$ & $\gamma$ &
 $\gamma$ \\
\noalign{\vskip3pt\hrule\vskip3pt}
$10^{16}$ &
 & \multicolumn{4}{c}{$f_\nu l_\nu\gtrsim3000$ Mpc for high URB, no EGMF$^a$}
 & $\nu\nu$ & $\gamma$ & $\gamma$ \\
 & \ref{fig:F1}, \ref{fig:F2} & high & any & SUSY & 10\% & $qq$
 & $N$ & $\gamma+N$, $N^c$ \\
 & \ref{fig:F1}, \ref{fig:F2} & high & $\lesssim10^{-9}$ & no-SUSY
 & 10\% & $qq$ & $\gamma$, $N^c$ & $\gamma$, $\gamma+N^d$ \\
 & & any & $\lesssim10^{-11}$ & any & $\lesssim10\%$ &
 $qq$, $ql$, $q\nu$ & & \\
 & & $\lesssim$ med & $\lesssim10^{-11}$ & -- & -- & $ll$, $l\nu$ &
 $\gamma$ & $\gamma$ \\
\end{tabular}
\smallskip
$^a$ viable for eV mass neutrinos if their overdensity $f_\nu$ over
a scale $l_\nu$ obeys specified condition for the high URB and vanishing
EGMF; for weaker URB/stronger EGMF condition relaxes/becomes more
stringent, respectively. \\
$^b$ dominant component of ``visible'' TD flux below and above GZK cutoff
at $\simeq70$ EeV.\\
$^c$ for EGMF $\gtrsim10^{-10}$ G.\\
$^d$ for EGMF $\gtrsim10^{-9}$ G.
\end{table}

We now present results from our flux calculations for a variety
of combinations of URBs, EGMFs, FFs, fractions $f_N$ of
nucleons created in quark fragmentation, and X particle decay
modes. Tab.~\ref{tab:T1}
identifies some of the scenarios that were found capable of
explaining HECRs at least above 100 EeV,
without violating any observational constraints, along with
the predicted composition of the TD component below and above
the GZK cutoff. The spectrum was normalized in the best
possible way to explain observed HECRs as being due either to
nucleon or $\gamma$-ray primaries. The flux below $\lesssim 20$ EeV
is presumably due to conventional acceleration and was not fit.
We remark that above $100\,$
EeV, the best fits for the viable scenarios from Tab.~\ref{tab:T1}
have acceptable likelihood significances (see Ref.~\cite{SLSB} for
details) and are consistent with the integral flux above
$300$ EeV estimated in Refs.~\cite{fe1,agasa1}, in contrast
to direct fits to the observed differential flux at $300$ EeV
\cite{PS} which would lead to an overproduction of the integral
flux at higher energies.

\begin{figure}[ht]
\centerline{\psfig{file=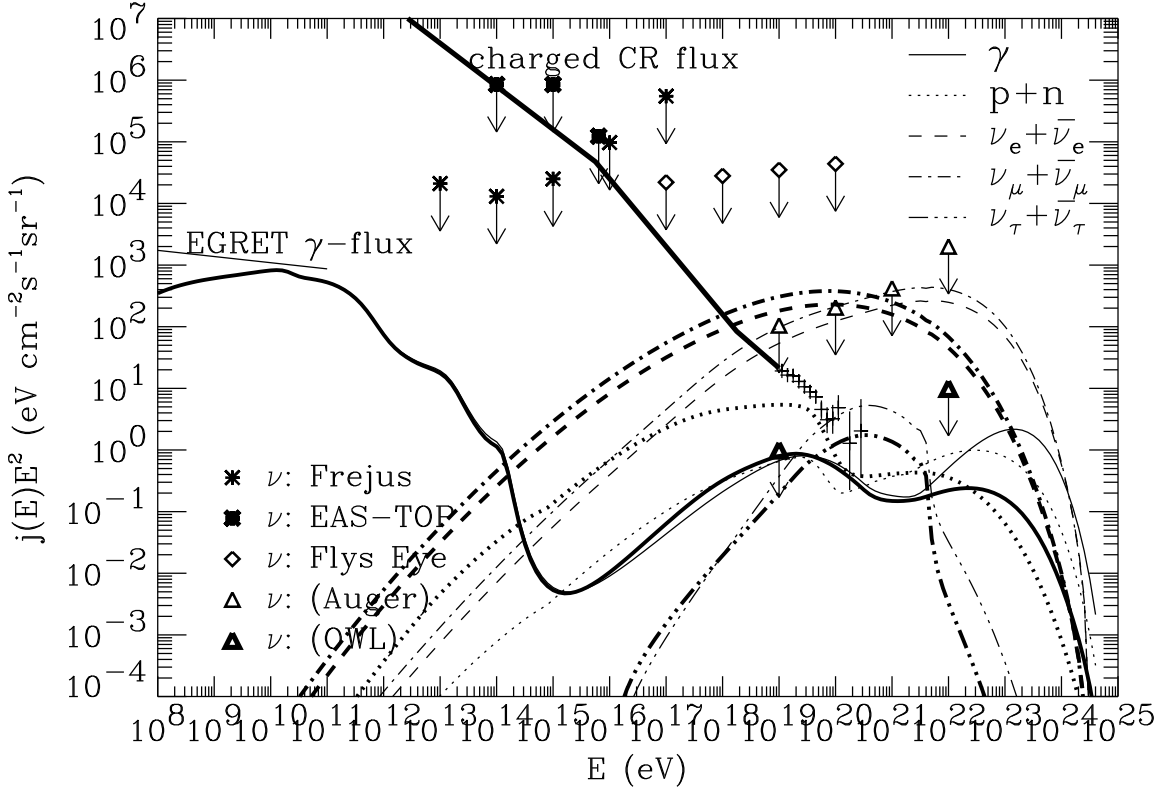,width=6in}}
\medskip
\caption[...]{Energy spectra of nucleons, $\gamma$-rays and neutrinos
for the TD model with $m_X=10^{16}$ GeV, $p=1$, and the decay
mode $X\to q+q$, assuming the
high URB version and an EGMF of $10^{-10}$ G. Thick
and thin lines represent the SUSY and no-SUSY
FFs, respectively. 1 sigma error bars are the combined data 
from the Haverah Park \cite{haverah}, Fly's Eye \cite{fe1} and
AGASA \cite{agasa1}
experiments above 10 EeV. Also shown are piecewise power
law fits to  the observed charged cosmic ray flux below 10 EeV, the
EGRET measurement of the diffuse $\gamma$-ray flux between
30 MeV and 100 GeV \cite{sreekumar}, and experimental neutrino flux
limits from Frejus \cite{rhode} and Fly's Eye \cite{baltrusaitis}, as
well as projected neutrino sensitivities of the future Pierre Auger
\cite{capelle} and NASA's OWL \cite{OWL} projects.
\label{fig:F1}}
\end{figure}

\begin{figure}[ht]
\centerline{\psfig{file=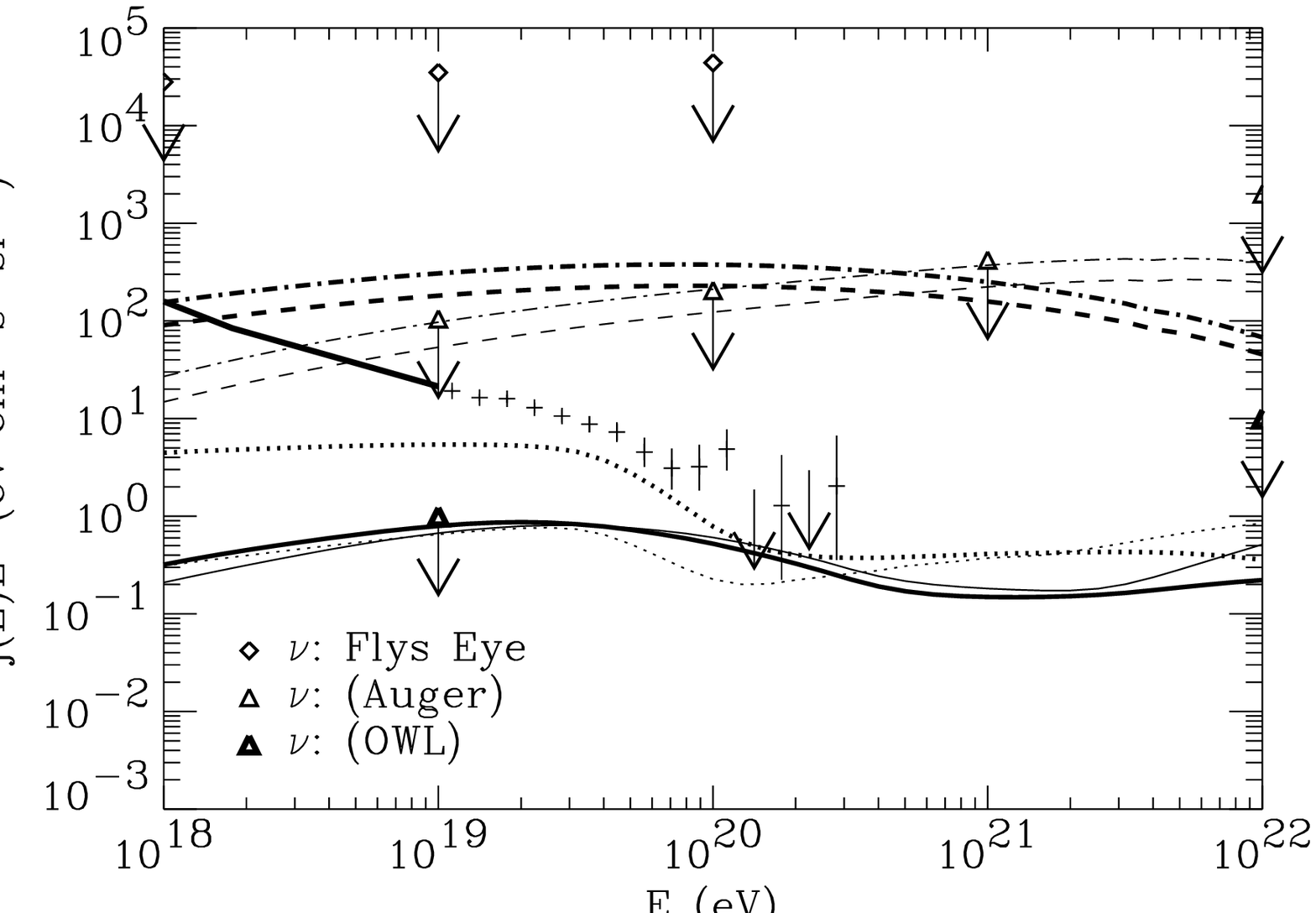,width=6in}}
\medskip
\caption[...]{Blow up of Fig.~\ref{fig:F1} for the fluxes at energies
above 1 EeV. The tau neutrino fluxes were omitted for clarity.
\label{fig:F2}}
\end{figure}

\begin{figure}[ht]
\centerline{\psfig{file=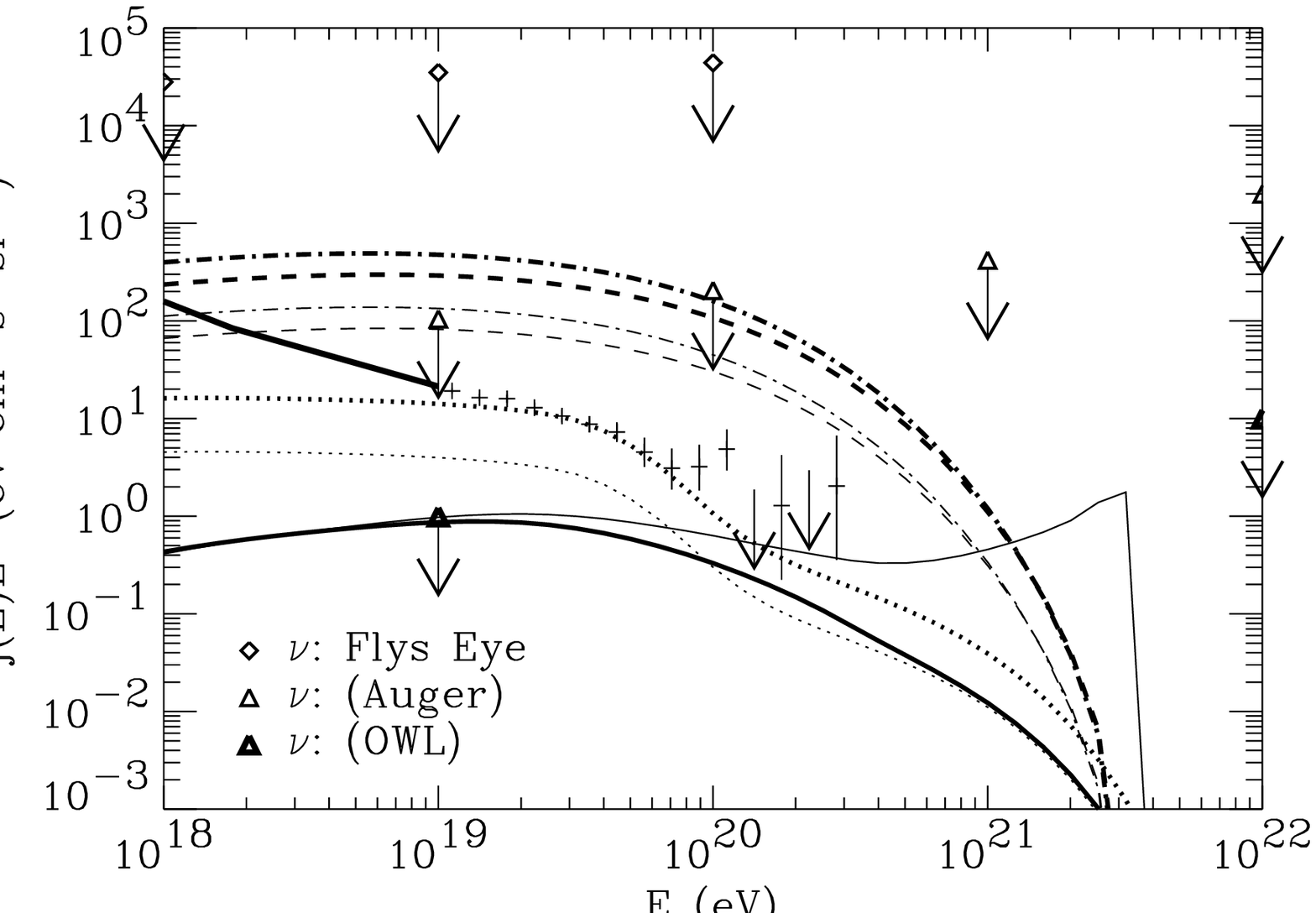,width=6in}}
\medskip
\caption[...]{Same as Fig.~\ref{fig:F2}, but for $m_X=10^{13}$ GeV,
and the no-SUSY FF, assuming a vanishing EGMF. Here,
the thick and thin lines represent the decay modes $X\to q+q$
and $X\to l+q$, respectively. The same normalization of the
GeV $\gamma$-ray flux as in Figs.~\ref{fig:F1},~\ref{fig:F2}
was used.
\label{fig:F3}}
\end{figure}

\begin{figure}[ht]
\centerline{\psfig{file=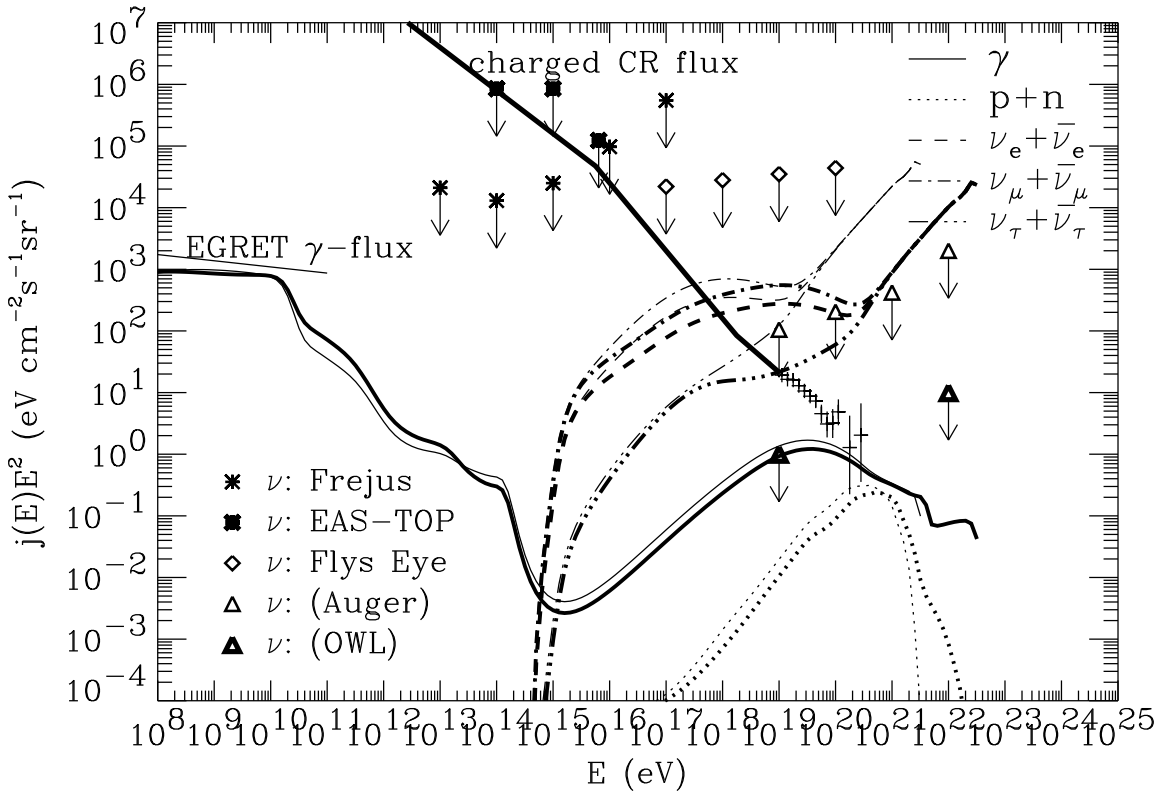,width=6in}}
\medskip
\caption[...]{Same as Fig.~\ref{fig:F1}, but for the pure
neutrino decay mode with no EGMF. Shown are the maximal UHE neutrino
fluxes allowed by the EGRET limit for $m_X=10^{14}$ GeV
(thick lines) and $m_X=10^{13}$ GeV (thin lines). For neutrino
clustering the lower limits from Tab.~\ref{tab:T1}, required
to explain HECRs, were assumed. This would correspond to
overdensities of $\simeq30$ and $\simeq75$ over a scale
$l_\nu\simeq5$ Mpc.
\label{fig:F4}}
\end{figure}

Figs.~\ref{fig:F1}--\ref{fig:F4} show the fluxes of some scenarios
indicated in Tab.~\ref{tab:T1}, along with current observational
constraints on and projected sensitivities of some future
experiments to $\gamma$-ray and neutrino fluxes. This demonstrates
consistency with present constraints within the normalization
ambiguity. In particular, EM energy injected at high
redshifts is recycled by cascading to lower energies, as can
be seen in Fig.~\ref{fig:F1}. TD models
are therefore significantly constrained~\cite{Chi,SJSB} by current
limits on the diffuse $\gamma$-ray background between
30 MeV and 100 GeV~\cite{sreekumar} which acts as a
``calorimeter'' and requires $Q^0_{\rm EM}\lesssim2.2
\times10^{-23}h(3p-1)\,{\rm eV}\,{\rm cm}^{-3}\,{\rm sec}^{-1}$
for the total energy injected into the EM channel. On the other
hand, it is not clear whether the observed diffuse background above
10 GeV can be fully accounted for by conventional sources
such as unresolved blazars~\cite{SS} and it has been suggested
that decays of heavy particles may provide a significant
contribution in this energy range~\cite{BSS}. As can be
seen in the figures, this is also the case for the TD
scenarios studied here. In these scenarios, the CMB depletes
the photon flux above 100 TeV, and the IR/O background
in the range 100 GeV--100TeV, recycling it to
energies below 100 GeV (see Fig.~\ref{fig:F1}).
The resulting background is {\it not} very sensitive to
the specific IR/O background model, however~\cite{AC}.
Constraints from limits on CMB distortions and light
element abundances from $^4$He-photodisintegration are
comparable to the bound from the directly observed
$\gamma$-rays~\cite{SJSB}.

Figs.~\ref{fig:F2} and~\ref{fig:F3} compare the UHE fluxes
from four TD scenarios indicated in Tab.~\ref{tab:T1}.
Fig.~\ref{fig:F2} compares the SUSY and
no-SUSY FF for $m_X=10^{16}$ GeV and is just the UHE part
of Fig.~\ref{fig:F1}. Fig.~\ref{fig:F3} compares
the two decay channels $X\to q+l$ and $X\to q+q$ for $m_X=10^{13}$ GeV,
assuming the no-SUSY FF. Both figures assume the high URB, an
EGMF of $10^{-10}$ G and $\lesssim10^{-11}$ G, respectively, and
a fraction $f_N\simeq10\%$ of nucleons created in quark fragmentation.
The present energy injection rate $Q^0_{\rm HECR}$ required to
produce the UHE fluxes $j_a(E)$ can be estimated as
\begin{equation}
  Q^0_{\rm HECR}\simeq10^{-22}\left(\frac{E^2 j_a(E)}{{\rm eV}\,
  {\rm cm}^{-2}\,{\rm sr}^{-1}\,{\rm s}^{-1}}\right)
  \left(\frac{x^2dN_a/dx}{0.004}\right)^{-1}
  \left(\frac{\lambda_a(E)}{10\,{\rm Mpc}}\right)^{-1}
  \,{\rm eV}\,{\rm cm}^{-3}\,{\rm s}^{-1}
  \,,\label{Q0}
\end{equation}
where $x=2E/m_X$, $\lambda_a(E)$ is the
effective attenuation length of species $a$, and the fiducial
values are for $E=100$ EeV, and the SUSY FF for $m_X=10^{16}$ GeV.
For the SUSY and no-SUSY FF, $Q^0_{\rm HECR}$ turns out to be minimal
around $m_X\sim10^{15}$ GeV and $10^{14}$ GeV, respectively, and
increases below and above that. This is confirmed by
the numerical calculations, as can be seen from
Figs.~\ref{fig:F2} and~\ref{fig:F3} and from Tab.~\ref{tab:T1}.
We therefore conclude that for most combinations of the URB and
the EGMF, the most poorly known astrophysical ingredients,
one can find combinations of possible decay modes and FFs that make
$p=1$ TD models with homogeneous source distribution
viable HECR explanations for $10^{13}\,{\rm GeV}\lesssim m_X
\lesssim10^{16}$ GeV. We note in this context that in some GUT
models, certain baryon number violating decay modes involving
leptons and quarks may violate limits on proton decay if $m_X$ is
too far below $10^{15}$ GeV, and may therefore be disfavored,
see, for example Ref.~\cite{AN}.

The energy loss and absorption lengths for UHE nucleons and photons
are short ($\lesssim 100$ Mpc). Thus, their predicted UHE fluxes are
independent of cosmological evolution. The $\gamma$-ray flux
below $\simeq 10^{11}$ eV, however, scales as the
total X particle energy release integrated over all redshifts
and increases with decreasing $p$~\cite{SJSB} roughly as
$1/(3p-1)$.
Scenarios with $p<1$ are therefore in general
ruled out (see Figs.~\ref{fig:F1} and ~\ref{fig:F2}), whereas
constant comoving injection rates ($p=2$) are well within the
limits. Since the EM flux above $\simeq10^{22}$ eV is
efficiently recycled to lower energies, the constraint on $p$
is in general less sensitive to $m_X$ then expected from
earlier CEL-based analytical estimates~\cite{Chi,SJSB}.

A specific $p=2$ scenario is realized in the case where the
supermassive X particles have a lifetime longer than the age of
the Universe and constitute part of cold dark matter, for which
non-thermal production in the early Universe has recently
been identified as a serious possibility~\cite{CKR}. In this case,
local clustering of the
sources in the galactic halo has to be taken into account
which provides the dominant contribution to observable
fluxes~\cite{BKV}. As a consequence, predicted spectra and
composition just reflect the injection spectrum, and the
diffuse $\gamma$-ray background at EGRET energies is not a
serious constraint.

We now turn to signatures of TD models at UHEs.
The full cascade calculations predict
$\gamma$-ray fluxes below 100 EeV that are a factor $\simeq3$
and $\simeq10$ higher than those obtained
using the CEL or absorption approximation often used in the
literature~\cite{BBV}, in the case of strong and weak URB,
respectively. This is also reflected by comparing Eq.~(\ref{Q0})
for the $\gamma$-ray flux with the energy injection rate
$Q^0_{\rm EM}$ allowed by the EGRET
observations, which yields $\lambda_\gamma\simeq100$ Mpc.
Again, this shows the importance
of non-leading particles in the development of unsaturated EM
cascades at energies below $\sim10^{22}$ eV.
As a consequence, in all viable HECR explaining cases with
only quarks among the X particle decay products,
we obtain $\gamma$/nucleon ratios above 200 EeV that are
$\gtrsim0.1/f_N$ for $m_X\gtrsim10^{15}$ GeV, and
about a factor 2 smaller for $m_X\lesssim10^{14}$ GeV,
even for the maximal URB, if the EGMF is $\lesssim10^{-11}$ G.
This ratio is about a factor 3 higher for the decay modes
containing a charged lepton.
Although a $\gamma$-ray primary for the HECR events is somewhat
disfavored currently~\cite{HVSV}, the compositional issue
is not settled yet, but future experiments such as the
Pierre Auger project~\cite{cronin} should be able to distinguish
$\gamma$-ray and nucleon primaries and test this signature.
We stress that there are viable scenarios with nucleon fluxes
that are comparable with or even higher than the $\gamma$-ray
flux at all energies in case of the high URB and/or for
a strong EGMF, and $f_N\simeq10\%$, see Figs.~\ref{fig:F2}
and~\ref{fig:F3}, and Tab.~\ref{tab:T1}.
The predictions from the SUSY FF
in Fig.~\ref{fig:F2} even seems able to explain all
cosmic rays above $\simeq50$ EeV, including the dip around
100 EeV, as a cross-over from nucleon domination to an about
equal mixture of $\gamma$-rays and nucleons.
The low $m_X$, pure quark decay modes such as the one shown
in Fig.~\ref{fig:F3} may be able to explain
all cosmic rays above 10 EeV by nucleon primaries, but also
tend to produce a more rapid fall-off of fluxes beyond 100 EeV,
which constitutes another testable signature.
The $\gamma$/nucleon ratio above 100 EeV is
about a factor 5 and 10 higher in the medium and minimal URB,
respectively,  as compared to the strong URB case, and in general
decreases strongly with increasing EGMF $\gtrsim10^{-11}$ G.

As indicated in Tab.~\ref{tab:T1}, another interesting scenario
involves the pure neutrino decay modes, also shown in Fig.~\ref{fig:F4}
for $m_X\leq10^{14}$ GeV. Here, the $\gamma$-rays and nucleons
are produced as secondaries from the interactions of these UHE
neutrinos with the relic neutrinos. Because $\gamma$-rays and
nucleons above 100 EeV must have been produced within a distance
$\lambda_a\simeq$
few Mpc from the observer, their flux is dominantly produced
by interactions with the locally clustered neutrinos if
$l_\nu f_\nu\gtrsim\lambda_a$. In this case, the energy fluence
in the secondaries is about $f_Z(f_\nu l_\nu/\lambda_Z)$
times the energy fluence in primary neutrinos around the Z resonance,
where $f_Z\simeq3\%$ is the fractional width of the Z and
$\lambda_Z\simeq38$ Gpc is the neutrino mean free path at the
Z resonance at zero redshift. In contrast, at energies where
the Universe is transparent for particles today, the dominant contribution
to their production by UHE neutrinos comes from
interactions with the unclustered relic neutrino component
at high redshift. This is because for energies $E\gtrsim E_{\rm res}$,
the probability for both resonant and non-resonant interaction
with the relic background per redshift interval is roughly
$(1+z)^{1/2}f_Z\,t_0/\lambda_Z$ in the matter dominated
regime, where $t_0$ is the age of the
Universe (for $E\lesssim E_{\rm res}$ this probability decreases
linearly with $E$). Because the Universe is opaque for $\gamma$-rays above
$\sim100$ TeV, this implies that the diffuse $\gamma$-ray background
below some energy $E$ is sensitive to the injection history
at $z\gtrsim(100\,{\rm TeV}/E)^{1/2}$. This explains why the
$\gamma$-ray background is steeper below 10 GeV than in the scenarios where
its dominant production is not by neutrino interactions, see
Fig.~\ref{fig:F4}. As a further consequence of neutrino interactions,
the secondary neutrino fluxes below an energy $E\lesssim E_{\rm res}$ are
sensitive to the injection history at $z\gtrsim E_{\rm res}/E$.
For $p=1$ scenarios, all other
fluxes are insensitive to the injection history at $z\gtrsim100$.
Since we are mainly interested in neutrino fluxes above
10 EeV and $\gamma$-ray fluxes above 100 MeV, it was therefore
sufficient to integrate injection up to $z=10^3$ which also
approximately marks the transition to radiation domination.
In addition, for $p=1$, the scaling of neutrino interaction rates implies
that the energy content in the secondaries, and thus in particular
in the low energy cascade $\gamma$-rays, constitutes a few percent
of the energy in UHE neutrinos. This fixes the maximally
allowed UHE neutrino flux which is shown in Fig.~\ref{fig:F4}
and implies the lower limit on $l_\nu f_\nu$ given in
Tab.~\ref{tab:T1} which is required if secondaries of UHE
neutrino interactions are to explain HECRs. The maximal energy
injection rate in neutrinos today allowed by the EGRET limit
is correspondingly higher than the upper bound on
$Q^0_{\rm EM}$ by about a factor 100. Observational
consequences of the UHE neutrino fluxes are discussed in the
following section.

The spectra predicted by scenarios where the X particles
decay into more than two quanta are qualitatively similar
to the ones for decay into two particles of the same type.
The details, however, depend on the energy distributions
of the decay products. To avoid introducing further model
dependent parameters, we do not consider such refinements
in the present paper as we do not consider scenarios where
the X particles themselves are created with relativistic
energies.

\section{Neutrino Flux Detection}
In order to discuss the prospects of detectability of
neutrino fluxes in TD scenarios we express the (in general energy
dependent) experimental sensitivities in terms of the
ice or water equivalent acceptance $A(E)$ (in units of volume
times solid angle). Future neutrino telescopes
of kilometer scale or larger will utilize the
detection of Cherenkov radiation from muons and EM showers
created in interactions of the neutrinos with nucleons either
in ice or in the deep sea.
Examples for experiments that aim at this
effective size are the ICECUBE version of the AMANDA experiment
at the South Pole, as well as the Radio Ice Cherenkov Experiment
(RICE) that aims at measuring the radio pulse from the neutrino
interaction, the French Astronomy with a Neutrino Telescope and
Abyss environmental RESearch (ANTARES) proposal, and the NESTOR
project in the Mediterranian.

An alternative method is to search for extensive air showers
initiated by electrons produced by neutrinos via the charged
current process. The interaction length of cosmic ray hadrons and
$\gamma$-rays is $\sim 100$ g cm$^{-2}$ above 10 EeV
and the probability of these strongly interacting particles initiating
air showers deeper than 1500 g cm$^{-2}$ is negligibly small.
Thus, showers starting deep in the atmosphere must be produced
by penetrating particles such as neutrinos. Large neutrino detectors
for measuring HECR air showers using the air fluorescence
technique, such as the High Resolution Fly's Eye now under
construction~\cite{hires} or the planned Japanese Telescope
Array~\cite{telarr} will have the potential
to search for deeply penetrating showers (DPS) initiated by
neutrinos~\cite{yoshida}. 
Their resolution of measurement of the atmospheric depth
at which the shower has its maximum particle density is expected to
be less than 30 g cm$^{-2}$ and the discrimination between DPS and
the regular air showers would be relatively straightforward. 
A possible contamination by a potential background of DPS,
secondary showers that result from
tau lepton decays deep in the atmosphere or from $\gamma$-ray
bremsstrahlung by muons, has been estimated to be less than $10^{-3}$
for 10 years observation by a typical fluorescence detector.
Hence UHE neutrino astronomy with air fluorescence detectors
is not background limited~\cite{yoshida}.

In addition, a giant surface array such as the proposed
Pierre Auger project~\cite{capelle} also has significant
sensitivity for neutrino detection by search for horizontal air showers.
The recently proposed satellite observatory
concept for an Orbiting Wide-angle Light collector (OWL)~\cite{OWL}
would increase the sensitivity to horizontal air showers by
at least another order of magnitude.

\begin{figure}[ht]
\centerline{\psfig{file=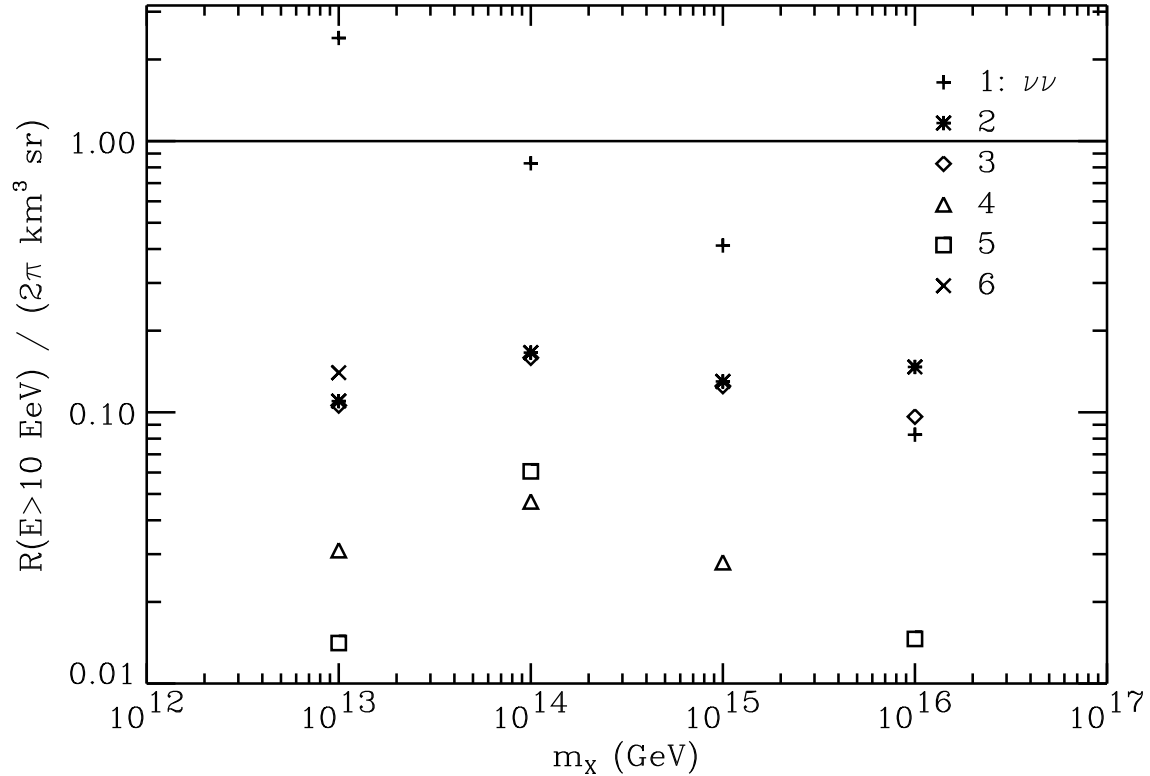,width=6in}}
\medskip
\caption[...]{Maximal event rates for muon neutrinos and anti-neutrinos
in a detector of $2\pi\,{\rm km}^3$ sr acceptance for the viable scenarios
from Tab.~\ref{tab:T1}, ordered by row number for given
$m_X$. Electron neutrino event rates are about a factor 2 smaller.
The rates for tau neutrinos are at least a factor 100 smaller still,
except if produced directly in the decay. The telescope array is
roughly sensitive to the range above the horizontal line, assuming
a duty cycle of 10 \% and a lifetime of 10 years.
\label{fig:F5}}
\end{figure}

Detection rates can be obtained by folding the
predicted fluxes with the product of the charged current
neutrino-nucleon cross section for
which we use the recent parametrization $\sigma_{\nu N}(E)
\simeq2.82\times10^{-32}(E/10\,{\rm EeV})^{0.402}\,{\rm cm}^2$
\cite{GQRS}, and the acceptance $A(E)$. Since the astrophysical
``background'' from other sources of UHE neutrinos, most notably
active galactic nuclei and Gamma Ray
Bursts~\cite{HZ,WB}, and the secondary neutrinos produced by
photopion production by HECR~\cite{yoshida}, is expected
to be negligible above 10 EeV, we present integral event
rates for neutrinos above 10 EeV in Fig~\ref{fig:F5} for
the viable HECR explaining TD models from Tab.~\ref{tab:T1}.
We furthermore assume an acceptance scaling as
$A(E)\propto E^{0.25}$ which seems to be implied by
experimental studies.

For a given $m_X$, the maximum of the neutrino event rates over
all decay modes except
the ones only involving neutrinos is typically reached for
the pure quark decay modes, except for $m_X=10^{13}$ GeV,
where the $l\nu$ mode produces the highest rates. As can be
seen from Fig.~\ref{fig:F5}, for all $m_X$ this maximum actually
saturates the general bound on the integral neutrino detection rate
$R(E)$ pointed out in Ref.~\cite{SLSC},
\begin{equation}
  R(E)\lesssim0.34\,r\left[{A(E)\over2\pi\,{\rm km}^3\,{\rm sr}}\right]
  \,\left({E\over10^{19}\,{\rm eV}}\right)^{-0.6}\,{\rm
  yr}^{-1}\,,\label{r2}
\end{equation}
for $E\gtrsim1$ PeV,
where $r$ is the ratio of energies injected into the neutrino
versus EM channel. This is not surprising because for all decay modes
except the ones only involving neutrinos, $r\leq0.5$.
The constraint Eq.~(\ref{r2}) is independent of the FF and arises
from comparing the energy content in neutrinos and $\gamma$-rays,
the latter being bounded from above by the EGRET measurement.

The highest possible rates are reached for the exclusive
neutrino decay mode at $m_X=10^{13}$ GeV for which the
bound Eq.~(\ref{r2}) is not applicable because $r=\infty$,
and the relevant quantity is the fraction of energy produced
as secondary $\gamma$-rays instead.
As can be seen from Fig.~\ref{fig:F4}, the neutrino flux
continues down to $\sim10^{15}$ eV in these scenarios
and can be comparable to fluxes predicted by models of
active galactic nuclei~\cite{HZ,WB}.
The maximally possible event rates from muon neutrinos above
1 PeV per year in a $2\pi\,{\rm km}^3\,{\rm sr}$ detector are
$\simeq5.5$ for $m_X=10^{13}$ GeV, and $\simeq3.5$ for
$m_X=10^{14}$ GeV.

In general, we conclude that at least the highest rates
predicted by TD models
should be observable by next generation experiments such
as the Pierre Auger Observatory and especially the OWL
project, as can also be seen from the sensitivities shown
in the figures.

\section{Summary}
Apart from the decay spectra and rates, the uncertainty of
flux predictions in TD scenarios is governed by astrophysical
uncertainties, mainly the universal radio background and the
large scale extragalactic magnetic field. Our calculations
show, however, that for most combinations of likely values
for these astrophysical parameters and the energy scale of
new physics, there are possible decay modes and fragmentation
functions that lead to scenarios explaining the highest
energy cosmic rays above the GZK cutoff, and some of them
even down to $\simeq10$ EeV, without violating observational
constraints on $\gamma$-ray and neutrino fluxes. For example,
an X particle of mass $m_X\simeq10^{16}$ GeV decaying into
quarks with a fragmentation function motivated by supersymmetry
can explain cosmic rays above $\simeq50$ EeV. This scenario
predicts a transition from a nucleon dominated component
to an about equal mixture of nucleons and $\gamma$-rays
above $\simeq100$ EeV in case of a relatively strong
universal radio background and a large scale magnetic field
$\lesssim10^{-10}$ G, a signature that should be
testable within the next few years. Other tests involve
GeV $\gamma$-rays whose flux comes close to the EGRET
measurement, and ultra high energy neutrino fluxes that
should be detectable by $\gtrsim$ few km scale neutrino
observatories which are now in the planning stage.

Another interesting viable class of scenarios involves pure
neutrino decay modes in the context of eV neutrino masses which
can yield even higher neutrino event rates up to a few per year
in km scale detectors above $\simeq10$ EeV for $m_X\lesssim10^{14}$
GeV. The neutrino flux extends down to $\sim1$ PeV in these
models where it can be comparable to predictions from
models of active galactic nuclei. Furthermore, for
a modest amount of clustering of neutrino dark matter on
the scale of the local Supercluster, secondary $\gamma$-ray
and nucleon production by neutrino interactions with the
clustered component can provide a significant fraction of the
highest energy cosmic ray flux.

\section*{Acknowledgements}
Special thanks go to the late David Schramm for his constant
encouragement and support for interdisciplinary research
in particle astrophysics.
We also thank Wolfgang Ochs, Jim Cronin, Chris Hill, and
Felix Aharonian for stimulating discussions, Paolo Coppi
for collaboration in earlier work, and Haim Goldberg for
helpful correspondence. P.B. is supported at NASA-GSFC by NAS/NRC
and NASA. At the University of Chicago this work was supported by DOE,
NSF and NASA.

\end{document}